\documentclass{article}
\usepackage{spconf,amsmath,graphicx}

\usepackage{enumitem}
\usepackage{multirow}
\setlist{nosep, leftmargin=14pt}

\usepackage{mwe} 

\title{Knowledge Distillation for Continual Learning of \\ Biomedical Neural Fields}
\name{Wouter Visser, Jelmer M. Wolterink}
\address{Department of Applied Mathematics, Technical Medical Center \\ 
University of Twente \\
Enschede, The Netherlands}
\begin{document}
%
\maketitle
\begin{abstract} 
Neural fields are increasingly used as a light-weight, continuous, and differentiable signal representation in (bio)medical imaging. However, unlike discrete signal representations such as voxel grids, neural fields cannot be easily extended. As neural fields are, in essence, neural networks, prior signals represented in a neural field will degrade when the model is presented with new data due to catastrophic forgetting. This work examines the extent to which different neural field approaches suffer from catastrophic forgetting and proposes a strategy to mitigate this issue. We consider the scenario in which data becomes available incrementally, with only the most recent data available for neural field fitting. In a series of experiments on cardiac cine MRI data, we demonstrate how knowledge distillation mitigates catastrophic forgetting when the spatiotemporal domain is enlarged or the dimensionality of the represented signal is increased. We find that the amount of catastrophic forgetting depends, to a large extent, on the neural fields model used, and that distillation could enable continual learning in neural fields. 
\end{abstract}
\begin{keywords}
Neural fields, continual learning, catastrophic forgetting, knowledge distillation
\end{keywords}
\section{Introduction}
\label{sec:intro}
Neural fields, also known as implicit neural representations, have emerged as an effective method for continuously representing signals on low-dimensional domains. 
A neural field represents a signal on a domain using a differentiable multilayer perceptron (MLP) that learns a mapping from domain coordinates to signal values at those coordinates. In (bio)medical imaging, neural fields are increasingly used for tasks such as image registration \cite{wolterink_implicit_2022} and shape analysis \cite{depaolisFastMedicalShape2025}, among others \cite{molaeiImplicitNeuralRepresentation2023}.

Neural fields are optimized to represent a given signal on a fixed domain. This signal may be known, in which case the neural field learns a direct mapping between coordinates and signal values, or unknown, requiring the solution of an inverse problem~\cite{reed_dynamic_2021, shen_nerp_2024}. Once optimized, a neural field can represent a fixed signal well, but its ability to incorporate new data is limited. In scenarios where data become available incrementally, such as longitudinal imaging or multimodal acquisitions, updating the neural field often degrades previously learned representations. This phenomenon, known as \textit{catastrophic forgetting}~\cite{mccloskey_catastrophic_1989}, reflects the trade-off between stability (retaining prior knowledge) and plasticity (learning new information).

While numerous strategies exist to mitigate catastrophic forgetting in conventional neural networks~\cite{wang_comprehensive_2024}, these methods do not directly translate to neural fields due to their unique characteristics.
First, neural fields operate on low-dimensional coordinate inputs from a limited domain, i.e., spatial and/or temporal coordinates. Theoretical research on continual learning shows that the similarity of the data contributes to forgetting~\cite{doan_theoretical_2021}, and thus, neural fields might be more susceptible to catastrophic forgetting. Second, neural field architectures typically account for spectral bias, i.e., the phenomenon in which a neural network struggles to represent high-frequency signals~\cite{rahaman_spectral_2019}. A common strategy is to change the activation function of the network~\cite{sitzmann_implicit_2020, liu_finer_2024}, but the choice of activation function also affects continual learning performance~\cite {goodfellow_empirical_2015}.

In this work, we address this open challenge with three contributions: (1) a systematic analysis of catastrophic forgetting in neural fields, (2) a memory-free knowledge distillation approach, and (3) an empirical validation on cardiac cine MRI.

\section{Materials and Methods}

\subsection{Neural Fields}
\label{sec:models}
A neural field is a neural network trained to represent a signal, with the signal's coordinates used as input and its values returned as output. Input coordinates $\mathbf{x}$ are selected from a domain, and target signal values $\mathbf{y}$ are either known or available as measurements produced by a forward operator. During training, a fitting loss $\mathcal{L}_{fit}$ is minimized between the model's output and the target signal using gradient descent.

Neural fields, like other neural networks, exhibit \textit{spectral bias}, a tendency to learn low-frequency signals faster than high-frequency signals~\cite{rahaman_spectral_2019}. This affects their ability to learn high-quality representations of biomedical images or shapes. Several strategies have been proposed to mitigate spectral bias. 

\subsubsection{Positional Encoding}
Positional encoding (P.E.) lifts input coordinates into a higher-dimensional space using predefined sine and cosine functions. These encoded coordinates are then used as input to a regular MLP with ReLU activations, which we refer to as the P.E. ReLU architecture. We use the encoding scheme proposed in~\cite{mildenhall_nerf_2021}, with  $L$ levels of increasing frequency:
\begin{align*}
    \gamma(x) = [&sin(2^0\pi x), cos(2^0\pi x), \dots\\
    &sin(2^{L-1}\pi x), cos(2^{L-1}\pi x)]
\end{align*}

\subsubsection{Activation Functions}
Spectral bias can also be overcome by replacing the commonly used ReLU with an alternative activation function. A common strategy is the use of sine activation functions in sinusoidal representation networks (SIREN)~\cite{sitzmann_implicit_2020}. The input to this activation function is scaled by a factor $\omega_0$, i.e., $\alpha(z)=sin(\omega_0 z)$.

One drawback of SIREN is the bias towards a particular subset of frequencies. Flexible spectral-bias tuning in Implicit Neural Representation (FINER)~\cite{liu_finer_2024} adjusts the SIREN activation function to $\alpha(z)=sin(\omega_0(|z|+1)z)$ to overcome this. Here, the frequency of $\alpha(z)$ increases with $|z|$.

\subsubsection{DINER}
\label{sec:DINER}

DINER \cite{zhu_disorder-invariant_2024} introduces a learned coordinate transformation via a hash table that maps input coordinates to an $L$-dimensional latent space. Points from this latent space are used as input to an MLP. The hash table and neural network parameters are jointly optimized so that the MLP only needs to learn a low-frequency signal while the hash table handles high-frequency variations.

\subsection{Knowledge Distillation for Neural Fields}
We consider a continual learning scenario where data arrives incrementally, causing either the \textit{input domain} or \textit{output signal} to expand. The neural field is optimized only on the most recent task $s$, but must retain knowledge from previous tasks $1:s-1$. 

\begin{figure}[t]
    \centering
    \includegraphics[width=0.7\linewidth]{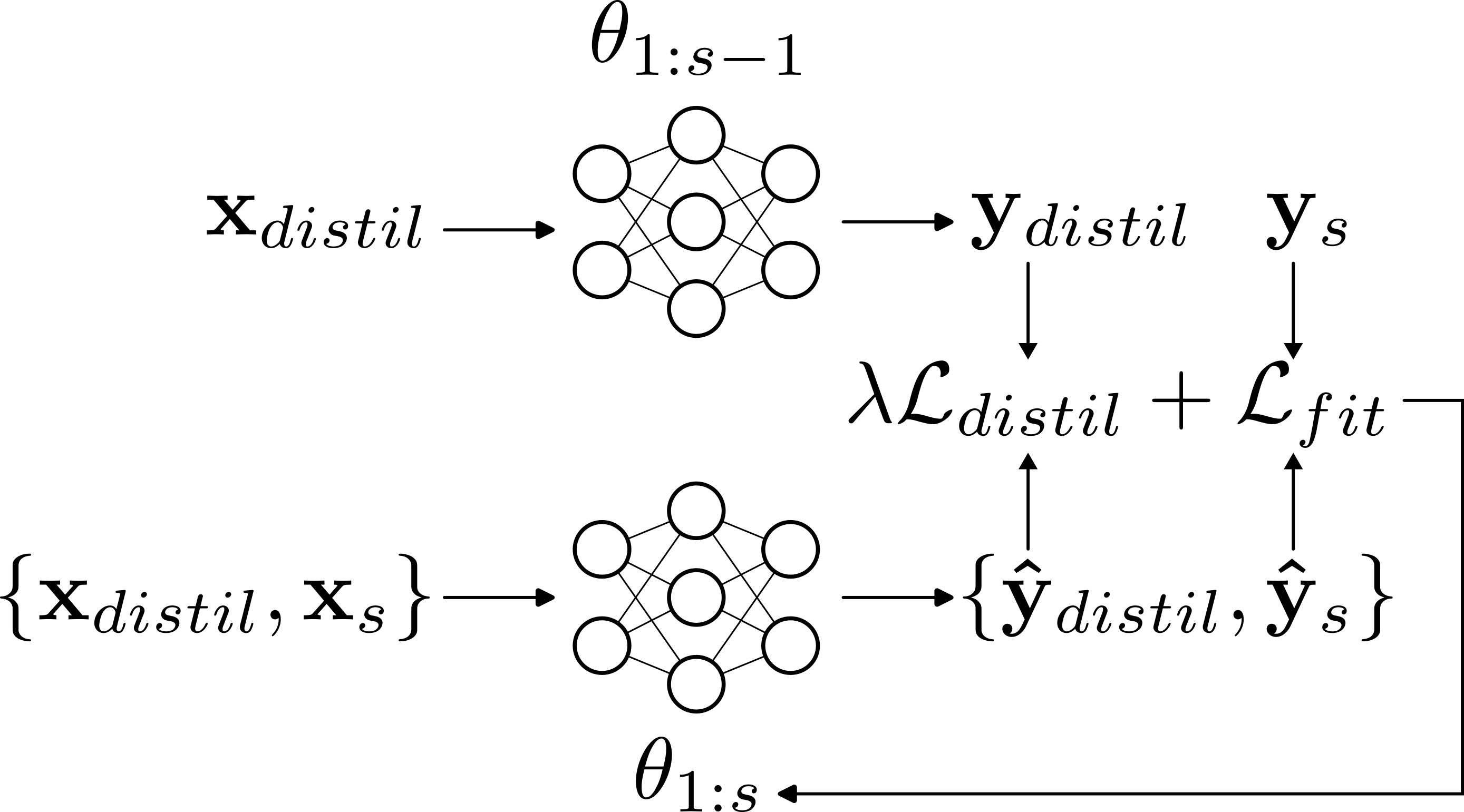}
    \caption{Schematic representation of knowledge distillation. $\theta_{1:s}$ denotes the parameters of the current model, while $\theta_{1:s-1}$ denotes the parameters of the previously trained model.}
    \label{fig:distillation_schematic}
\end{figure}

 To address this, we propose a memory-free knowledge distillation strategy (Fig. \ref{fig:distillation_schematic}).  Let the model parameters after learning tasks $1:s-1$ be $\theta_{1:s-1}$. We proceed as follows to learn the signal $s$ additionally. As in normal neural field training, we iteratively optimize the parameters $\theta_{1:s}$ by sampling coordinates $\mathbf{x}_s$ and their associated signals $\mathbf{y}_s$ from the new task $s$ and minimizing the fitting loss $\mathcal{L}_{fit}$. In addition, we sample a set of coordinates $\mathbf{x}_{distil}$ from the domain of the previously learned tasks (which could be identical to the domain of $s$). This sampling is uniform, but could be adjusted to reflect important or complex parts of the signal. Then, an output signal $\mathbf{y}_{distil}$ is generated for $\mathbf{x}_{distil}$ using the model with its previous, frozen, parameter set $\theta_{1:s-1}$ as well as its current parameter set $\theta_{1:s}$, which we denote by $\mathbf{\hat{y}}_{distil}$. As the goal is for the new outputs to be consistent with the old outputs, we compute loss $\mathcal{L}_{distil}$, which is combined with the fitting loss $\mathcal{L}_{fit}$,
\begin{equation*}
\mathcal{L}_{total} = \mathcal{L}_{fit} + \lambda\mathcal{L}_{distil},
\end{equation*}
where $\lambda$ balances plasticity and stability. This approach requires no access to previous data, only the prior model parameters, making it highly memory-efficient. 

\section{Experiments and Results}
We evaluated the proposed approach on the training set of the Automated Cardiac Diagnosis Challenge (ACDC) challenge~\cite {bernard_deep_2018}, which contains 3D+t MRI scans of 100 patients. Each scan consists of a sequence of 3D frames covering one complete cardiac cycle. We normalized voxel intensities in each patient to the range $[0, 1]$. Ground truth segmentation masks for the left ventricle, myocardium, and right ventricle were available for end-diastolic and end-systolic frames

All neural networks were implemented as multi-layer perceptrons (MLPs) with three hidden layers, each containing 256 neurons. Hidden layers used ReLU activation functions, unless an alternative was used, as in SIREN or FINER. Output activation functions were linear for image reconstruction and softmax for segmentation mask reconstruction. As a fitting loss $\mathcal{L}_{fit}$, we used a Huber loss in case images were fitted, and a binary cross-entropy loss in case segmentation masks were fitted. All models were trained for 500 iterations for each task, using Adam with a learning rate of 0.01 for the DINER models and 0.001 for the other models. For the P.E. ReLU models, we used a positional encoding level $L=10$. The SIREN and FINER models used $\omega_0=15$ and $\omega_0=5$, respectively. The DINER models used latent space dimensionality $L=1$. 

\begin{figure}
    \centering
    \includegraphics[width=\linewidth]{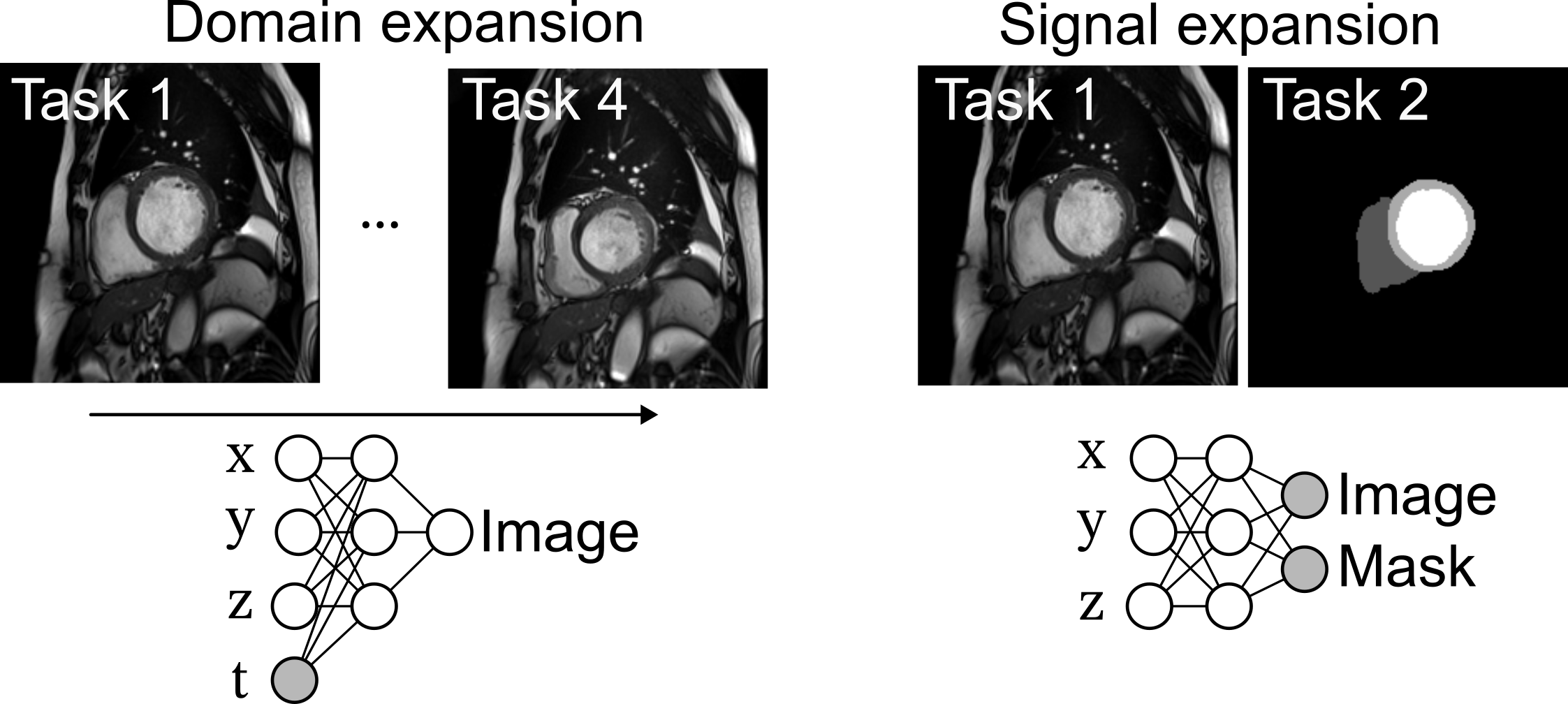}
    \caption{We consider domain expansion and signal expansion. Domain expansion affects the input values' range of one or more input coordinates, in this case, time $t$. Signal expansion affects the output values. In this case, we first fit an image, and then its corresponding segmentation mask.}
    \label{fig:placeholder}
\end{figure}

We considered two different continual learning scenarios (Fig. \ref{fig:placeholder}). In \textit{domain expansion}, we optimized a spatiotemporal neural field and expanded the domain along the time dimension. This could be the case, for example, if patient data is acquired longitudinally. In \textit{signal expansion}, we optimized a neural field and added a signal on the same domain. This could be the case if multiple aligned images are being acquired, or, as was the case here, a segmentation mask is added to an image. Models were trained incrementally, both without and with knowledge distillation. In our results, we refer to these optimization strategies as \textit{baseline} and \textit{distillation}. In all cases, we rescaled spatial and temporal coordinates to $[-1, 1]$. Image reconstruction quality was assessed using SSIM and PSNR metrics to evaluate global intensity and structural reconstruction, respectively. Segmentation quality was measured using the Dice similarity coefficient (DSC). All metrics were averages over 100 patients.

\begin{figure}[t]
    \centering
    \includegraphics[width=\linewidth]{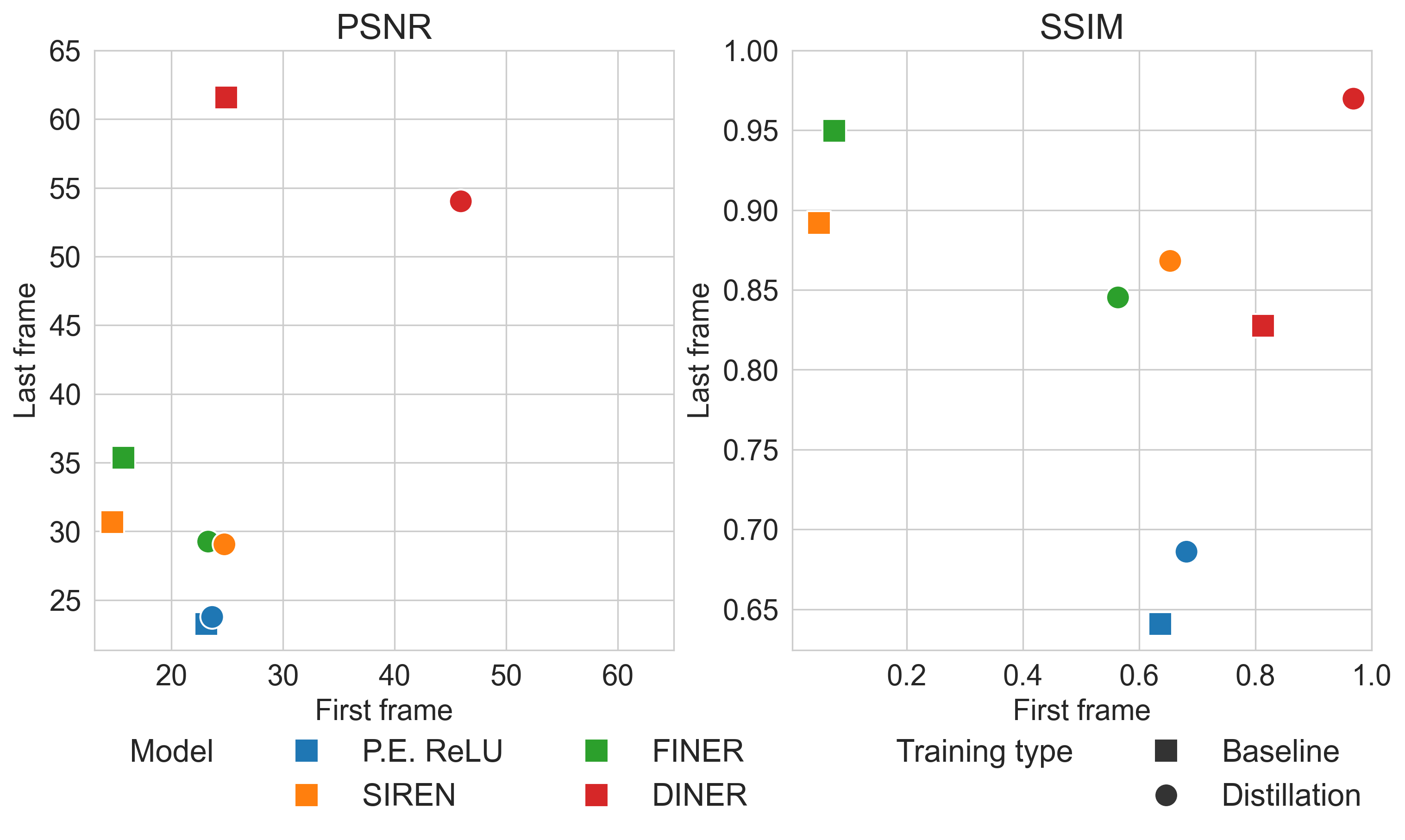}
    \caption{Comparison of reconstruction quality between the first and last frame of the 3D+time dataset in the domain expansion setting.}
    \label{fig:acdc_4d_graph}
\end{figure}

\begin{figure}[t]
    \centering
    \includegraphics[width=\linewidth]{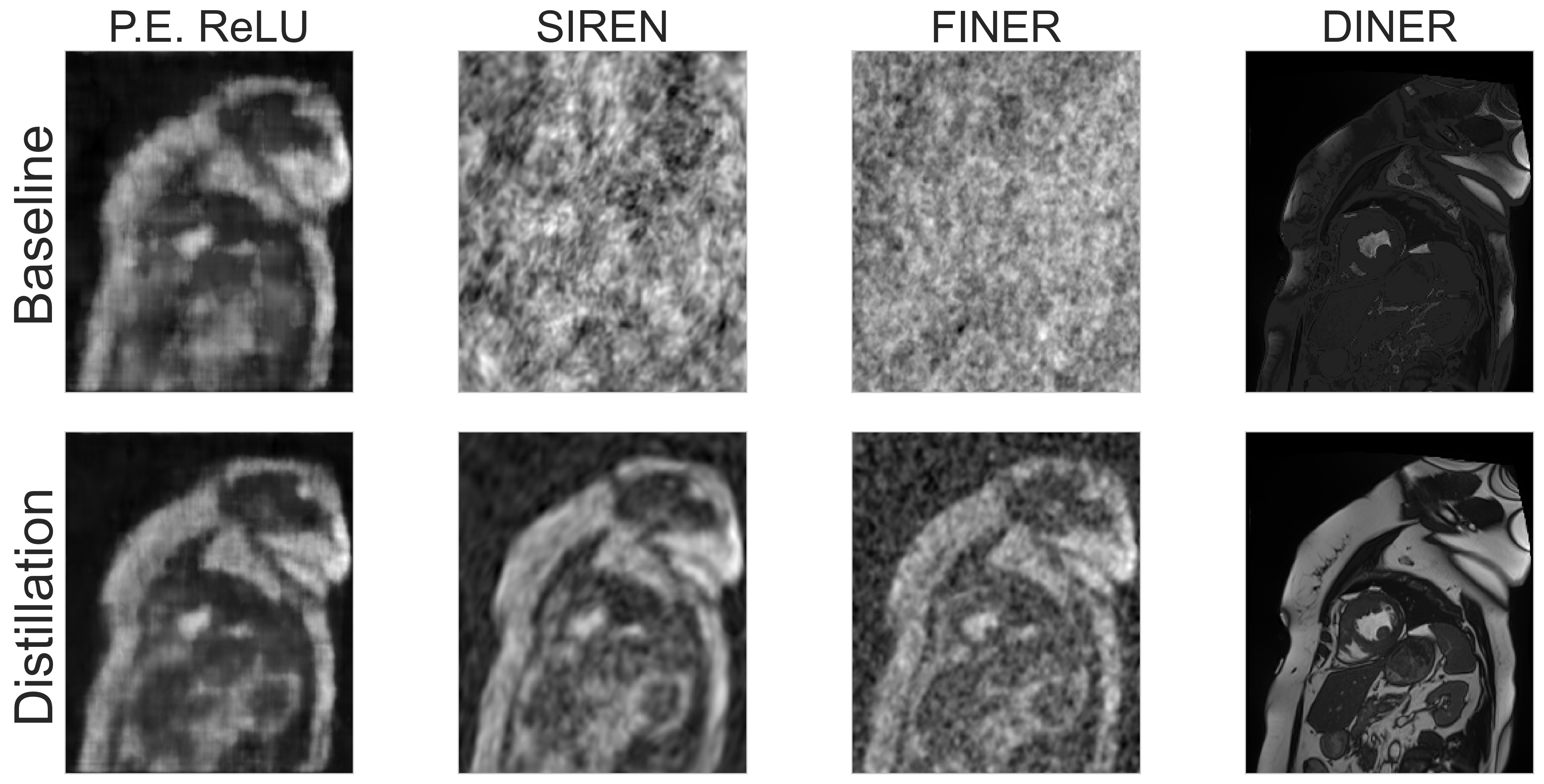}
    \caption{Examples of reconstruction of the first frame of the 3D+t dataset for models trained without and with distillation.}
    \label{fig:acdc_4d_examples}
\end{figure}

\subsection{Domain expansion}
From each 3D+t ACDC image sequence, we selected the first and last frames, along with two intermediate frames. A neural field was first fitted to the initial frame and then incrementally updated with the others at  $t= [-1, -\frac{1}{3}, \frac{1}{3}, 1]$. 

Figure \ref{fig:acdc_4d_graph} shows average SSIM and PSNR for the first ($x$-axis) and last ($y$-axis) frames when using the baseline or distillation strategies. Models closer to the top-right exhibit better stability (retaining the first frame) and plasticity (fitting the last frame). All models trained without distillation perform poorly on the first frame, suggesting forgetting, especially SIREN and FINER (see also Fig.~\ref{fig:acdc_4d_examples}). DINER behaved differently across metrics: SSIM remained high while PSNR dropped after learning multiple frames, suggesting structural preservation but intensity loss. This likely stems from its coordinate-wise encoding, which is not updated for previous frames, thereby maintaining their structure while intensity performance worsens.

The plot also shows that all models benefit from knowledge distillation to improve performance on early tasks, without detrimentally affecting performance on late tasks. This pattern is most prominent in the DINER model and least visible in the P.E. ReLU model.

\begin{figure}[t]
    \centering
    \includegraphics[width=\linewidth]{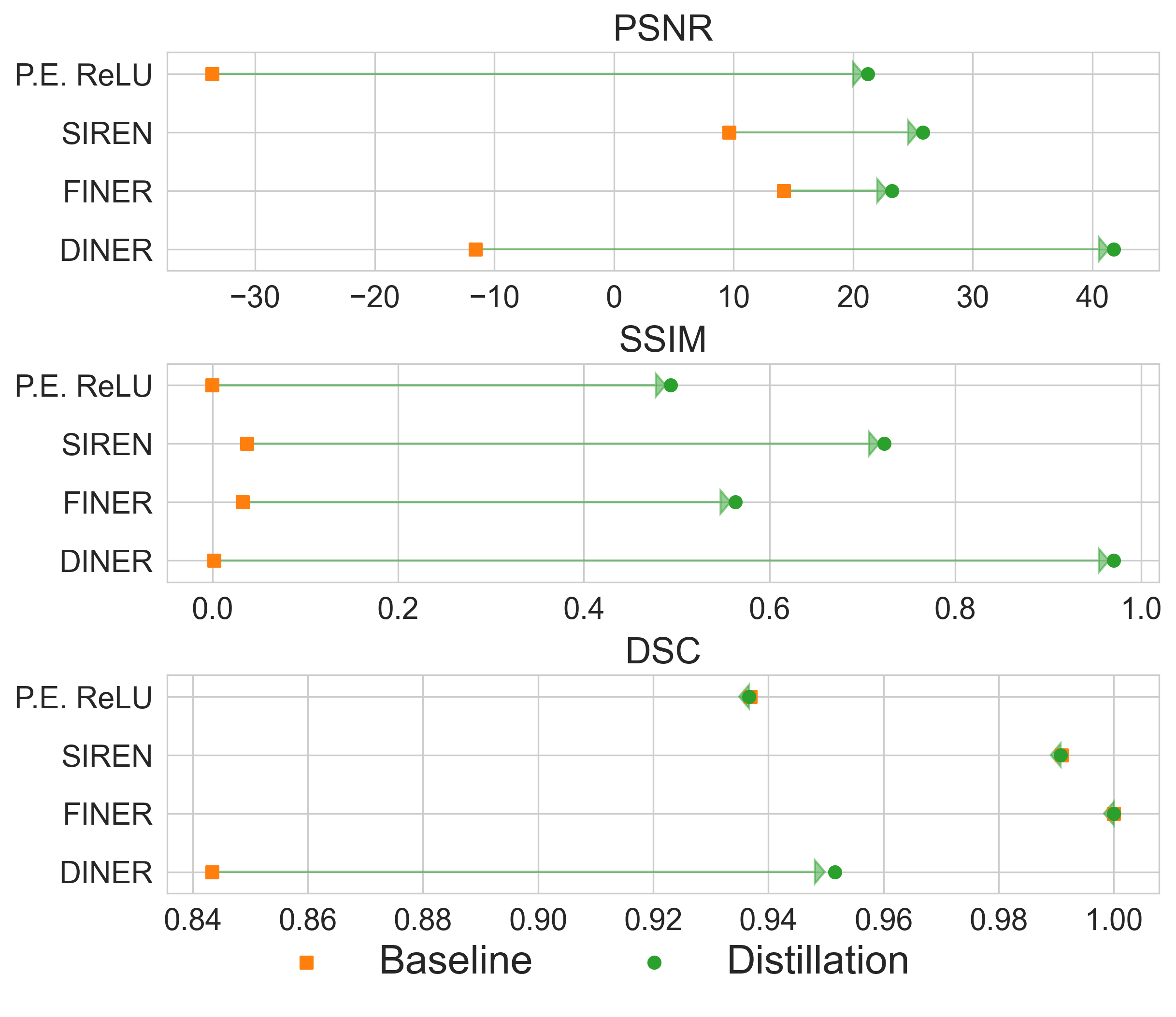}
    \caption{Comparison of reconstruction and segmentation performance of training types in the signal expansion setting.}
    \label{fig:acdc_seg_graph}
\end{figure}

\begin{figure}[t]
    \centering
    \includegraphics[width=\linewidth]{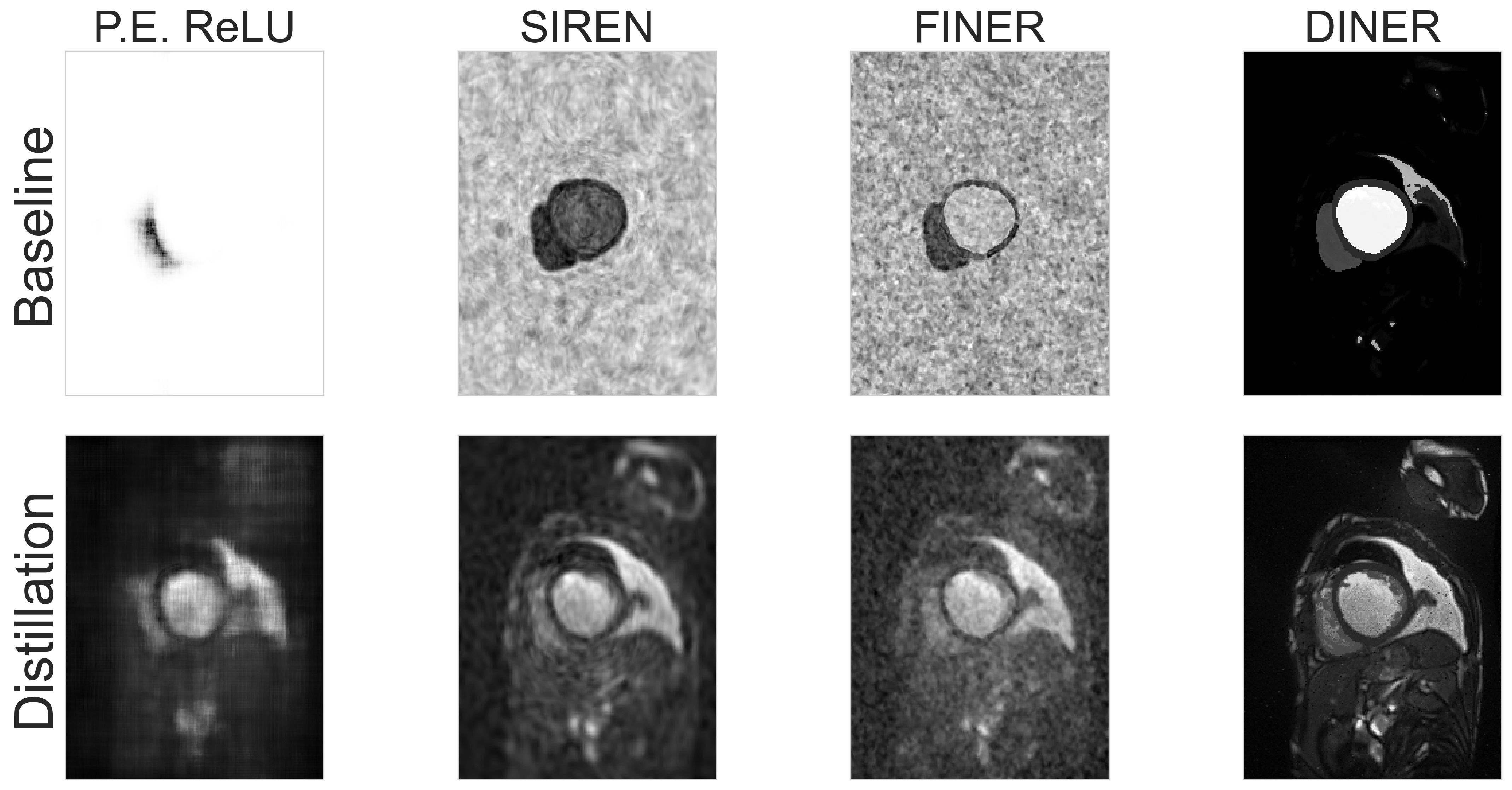}
    \caption{Examples of image reconstruction (Task 1) after learning segmentation masks (Task 2) for models trained without and with distillation}
    \label{fig:acdc_seg_examples}
\end{figure}

\subsection{Signal expansion}
We used 3D ACDC images of end-diastole and end-systole with corresponding segmentation masks. A neural field was first trained to represent the image, and then extended to also represent the aligned segmentation mask. For this, we added output nodes for a one-hot encoding of classes.

Figure \ref{fig:acdc_seg_graph} shows PSNR and SSIM values for the image task after learning segmentation masks. Unlike the domain expansion experiment, P.E. ReLU and DINER exhibited catastrophic forgetting. Distillation substantially improves performance on the MRI representation for all models, with the DINER model achieving the best results. Fig. \ref{fig:acdc_seg_examples} shows an example of an image frame reconstruction after the model has learned the segmentation task, for all four models and baseline and distillation approaches. This shows that all baseline models failed in different ways to reconstruct the image.

Figure \ref{fig:acdc_seg_graph} also shows DSC scores for the second task. We found that all models learned the segmentation task effectively, both with and without distillation, achieving DSC scores between 0.84 and 1.00. Lower DSC scores for the DINER baseline can be attributed to failure to represent the segmentation mask in individual patients. This issue was mitigated by adding distillation, suggesting that the inclusion of a (distilled) image reconstruction task could stabilize segmentation mask reconstruction. 

\section{Discussion}
Our results show that neural fields are highly susceptible to catastrophic forgetting in continual learning settings, both during domain and signal expansion. We found that, to a large extent, the ability of a neural field to incorporate new signals depends on the chosen strategy to overcome spectral bias. Moreover, we demonstrated that knowledge distillation effectively remedies catastrophic forgetting across all models. 

We used the same neural network architecture in all cases, but found that SIREN and FINER, both strategies that modify the activation function, were most susceptible to catastrophic forgetting. On the other hand, the P.E. ReLU approach did not suffer from catastrophic forgetting in the domain expansion experiment, but in general failed to achieve high-quality reconstructions. This is a new finding, as previous works on continual learning in non-biomedical neural fields have primarily considered P.E. ReLU models~\cite{chung_meil-nerf_2025, cai_clnerf_2023} 
We hypothesize that these differences can be partly explained by the neural tangent kernel (NTK) of neural networks, which is affected differently by strategies to overcome spectral bias. Alternatively, continual learning using SIREN and FINER might be sensitive to the choice of hyperparameters. In future work, we aim to gain a better theoretical understanding of the complex interplay between the NTK and continual learning. 

We found that DINER consistently outperformed alternatives for image reconstruction. In part, this may be due to the greater number of parameters in DINER compared to the other three models, which used only an MLP. These parameters are mainly stored in the hash table, which contains an entry per coordinate. In contrast, the MLP models had 5-10\% of the total number of voxels in the images, highlighting their potential as a data-efficient image representation.

We considered a distillation approach for memory-free continual learning and demonstrated its effectiveness, but various continual-learning approaches in neural networks have been proposed.  This includes methods that directly act on individual model weights, such as elastic weight consolidation (EWC)~\cite{kirkpatrick_overcoming_2017}. In preliminary experiments, we found that EWC was unable to facilitate continual learning, a finding corroborated in continual learning for NeRFs~\cite{cai_clnerf_2023, chung_meil-nerf_2025}. In future work, we will investigate alternative strategies, such as orthogonal gradient descent~\cite{doan_theoretical_2021} that address the similarity of the input domain across tasks. 

In conclusion, while neural fields are highly susceptible to catastrophic forgetting, distillation approaches might be a suitable way to mitigate this problem in continual learning for biomedical neural fields. 

\section{Compliance with Ethical Standards}
This research study was conducted retrospectively using human subject data made available in open access by Bernard et al.~\cite{bernard_deep_2018}. Ethical approval was not required as confirmed by the license attached with the open access data.

\section{Conflicts of interest}
No funding was received for conducting this study. The authors have no relevant financial or non-financial interests to disclose.

\bibliographystyle{IEEEbib}
\bibliography{strings,refs}

\begin{thebibliography}{10}

\bibitem{wolterink_implicit_2022}
Jelmer~M. Wolterink, Jesse~C. Zwienenberg, and Christoph Brune,
\newblock ``Implicit neural representations for deformable image registration,''
\newblock in {\em Proceedings of {The} 5th {International} {Conference} on {Medical} {Imaging} with {Deep} {Learning}}. Dec. 2022, pp. 1349--1359, PMLR,
\newblock ISSN: 2640-3498.

\bibitem{depaolisFastMedicalShape2025}
Gaia Romana De~Paolis et~al.,
\newblock ``Fast medical shape reconstruction via~meta-learned implicit neural representations,''
\newblock in {\em Shape in {{Medical Imaging}}}, Cham, 2025, pp. 189--204, Springer Nature Switzerland.

\bibitem{molaeiImplicitNeuralRepresentation2023}
Amirali~Molaei et~al.,
\newblock ``Implicit neural representation in medical imaging: A comparative survey,''
\newblock in {\em 2023 {{IEEE}}/{{CVF International Conference}} on {{Computer Vision Workshops}} ({{ICCVW}})}, Oct. 2023, pp. 2373--2383.

\bibitem{reed_dynamic_2021}
Albert W.~Reed et~al.,
\newblock ``Dynamic {CT} reconstruction from limited views with implicit neural representations and parametric motion fields,''
\newblock in {\em 2021 {IEEE}/{CVF} {International} {Conference} on {Computer} {Vision} ({ICCV})}, Oct. 2021, pp. 2238--2248,
\newblock ISSN: 2380-7504.

\bibitem{shen_nerp_2024}
Liyue Shen, John Pauly, and Lei Xing,
\newblock ``{NeRP}: implicit neural representation learning with prior embedding for sparsely sampled image reconstruction,''
\newblock {\em IEEE Transactions on Neural Networks and Learning Systems}, vol. 35, no. 1, pp. 770--782, Jan. 2024,
\newblock Conference Name: IEEE Transactions on Neural Networks and Learning Systems.

\bibitem{mccloskey_catastrophic_1989}
Michael McCloskey and Neal~J. Cohen,
\newblock ``Catastrophic interference in connectionist networks: the sequential learning problem,''
\newblock in {\em Psychology of {Learning} and {Motivation}}, Gordon~H. Bower, Ed., vol.~24, pp. 109--165. Academic Press, Jan. 1989.

\bibitem{wang_comprehensive_2024}
Liyuan Wang, Xingxing Zhang, Hang Su, and Jun Zhu,
\newblock ``A comprehensive survey of continual learning: theory, method and application,''
\newblock {\em IEEE Transactions on Pattern Analysis and Machine Intelligence}, vol. 46, no. 8, pp. 5362--5383, Aug. 2024,
\newblock Conference Name: IEEE Transactions on Pattern Analysis and Machine Intelligence.

\bibitem{doan_theoretical_2021}
Thang Doan, Mehdi~Abbana Bennani, Bogdan Mazoure, Guillaume Rabusseau, and Pierre Alquier,
\newblock ``A theoretical analysis of catastrophic forgetting through the {NTK} overlap matrix,''
\newblock in {\em Proceedings of the 24th {International} {Conference} on {Artificial} {Intelligence} and {Statistics}}. Mar. 2021, pp. 1072--1080, PMLR,
\newblock ISSN: 2640-3498.

\bibitem{rahaman_spectral_2019}
Nasim~Rahaman et~al.,
\newblock ``On the spectral bias of neural networks,''
\newblock in {\em Proceedings of the 36th {International} {Conference} on {Machine} {Learning}}. May 2019, pp. 5301--5310, PMLR,
\newblock ISSN: 2640-3498 shortConferenceName: ICML.

\bibitem{sitzmann_implicit_2020}
Vincent Sitzmann, Julien Martel, Alexander Bergman, David Lindell, and Gordon Wetzstein,
\newblock ``Implicit {Neural} {Representations} with {Periodic} {Activation} {Functions},''
\newblock in {\em Advances in {Neural} {Information} {Processing} {Systems}}. 2020, vol.~33, pp. 7462--7473, Curran Associates, Inc.

\bibitem{liu_finer_2024}
Zhen~Liu et~al.,
\newblock ``{FINER}: flexible spectral-bias tuning in implicit {NEural} representation by variableperiodic activation functions,''
\newblock in {\em 2024 {IEEE}/{CVF} {Conference} on {Computer} {Vision} and {Pattern} {Recognition} ({CVPR})}, June 2024, pp. 2713--2722,
\newblock ISSN: 2575-7075.

\bibitem{goodfellow_empirical_2015}
Ian~J. Goodfellow, Mehdi Mirza, Da~Xiao, Aaron Courville, and Yoshua Bengio,
\newblock ``An empirical investigation of catastrophic forgetting in gradient-based neural networks,'' Mar. 2015,
\newblock arXiv:1312.6211 [stat].

\bibitem{mildenhall_nerf_2021}
Ben Mildenhall, Pratul~P. Srinivasan, Matthew Tancik, Jonathan~T. Barron, Ravi Ramamoorthi, and Ren Ng,
\newblock ``{NeRF}: representing scenes as neural radiance fields for view synthesis,''
\newblock {\em Commun. ACM}, vol. 65, no. 1, pp. 99--106, Dec. 2021.

\bibitem{zhu_disorder-invariant_2024}
Hao~Zhu et~al.,
\newblock ``Disorder-invariant implicit neural representation,''
\newblock {\em IEEE Transactions on Pattern Analysis and Machine Intelligence}, vol. 46, no. 8, pp. 5463--5478, Aug. 2024.

\bibitem{bernard_deep_2018}
Olivier~Bernard et~al.,
\newblock ``Deep learning techniques for automatic {MRI} cardiac multi-structures segmentation and diagnosis: is the problem solved?,''
\newblock {\em IEEE Trans Med Imag}, vol. 37, no. 11, pp. 2514--2525, Nov. 2018.

\bibitem{chung_meil-nerf_2025}
Jaeyoung Chung, Kanggeon Lee, Sungyong Baik, and Kyoung~Mu Lee,
\newblock ``{MEIL}-{NeRF}: memory-efficient incremental learning of neural radiance fields,''
\newblock {\em IEEE Access}, pp. 1--1, 2025.

\bibitem{cai_clnerf_2023}
Zhipeng Cai and Matthias Müller,
\newblock ``{CLNeRF}: continual learning meets {NeRF},''
\newblock in {\em 2023 {IEEE}/{CVF} {International} {Conference} on {Computer} {Vision} ({ICCV})}, Oct. 2023, pp. 23128--23137,
\newblock ISSN: 2380-7504.

\bibitem{kirkpatrick_overcoming_2017}
James~Kirkpatrick et. al.,
\newblock ``Overcoming catastrophic forgetting in neural networks,''
\newblock {\em Proceedings of the National Academy of Sciences}, vol. 114, no. 13, pp. 3521--3526, Mar. 2017,
\newblock Publisher: Proceedings of the National Academy of Sciences.

\end{thebibliography}

\end{document}